\documentclass[proof]{WileyASNA-v1}

\articletype{ORIGINAL ARTICLE}

\received{March 24, 2018}
\revised{May 7, 2018}
\accepted{May 9, 2018}

\usepackage[rightcaption]{sidecap}
\usepackage{nicefrac}
\usepackage{hyperref}
\hypersetup{
    final=true,
    pageanchor=true,
    colorlinks=true,
    breaklinks=true,
    linkcolor=blue,
    citecolor=blue,
    urlcolor=blue,
    pdfpagemode=UseNone,
    pdftitle={On the extent of the moat flow in axisymmetric sunspots},
    pdfauthor={M. Verma et al.},
    pdfsubject={Solar Physics},
    pdfkeywords={sunspots, activity, photosphere, data analysis, 
        image processing}}

\newcommand\phn{\phantom{0}}

\definecolor{myGreen}{cmyk}{0.992,0.,0.083, 0.525}

\definecolor{myDarkRed}{rgb}{0.698, 0.094, 0.133}

\newcommand{\ns}{\hspace*{-5pt}}

\begin{document}


\title{On the extent of the moat flow in axisymmetric sunspots}

\author[1]{M.\ Verma*}
\author[1,2]{P.\ Kummerow}
\author[1]{C.\ Denker}

\authormark{M.\ Verma \textsc{et al.}}

\address[1]{\orgname{Leibniz-Institut f\"{u}r Astrophysik Potsdam (AIP)}, 
    \orgaddress{\state{Potsdam}, \country{Germany}}}

\address[2]{\orgdiv{Universit{\"a}t Potsdam}, 
    \orgname{Institut f{\"u}r Physik und Astronomie}, 
    \orgaddress{\state{Potsdam}, \country{Germany}}}

\corres{*\email{mverma@aip.de}}

\presentaddress{M.\ Verma, Leibniz-Institut f\"{u}r Astrophysik Potsdam (AIP),
    An der Sternwarte~16,
    14482 Potsdam,
    Germany}

\abstract[Abstract]{Unipolar, axisymmetric sunspots are figuratively called 
``theoretician's sunspots'' because their simplicity supposedly makes them 
more suitable for theoretical descriptions or numerical models. On 2013
November~18, a very large specimen (active region NOAA~11899) crossed the
central meridian of the Sun. The moat flow associated with this very large 
spot is quantitatively compared to that of a medium and a small sunspot to determine the 
extend of the moat flow in different environments. We employ continuum images
and magnetograms of the Helioseismic and Magnetic Imager (HMI) as well as
extreme ultraviolet (EUV) images at $\lambda$160~nm of the Atmospheric 
Imaging Assembly (AIA), both on board the Solar Dynamics Observatory (SDO), 
to measure horizontal proper motions with Local Correlation Tracking (LCT) 
and flux transport velocities with the Differential Affine Velocity Estimator
(DAVE). We compute time-averaged flow maps ($\pm$6~hours around meridian
passage) and radial averages of photometric, magnetic and flow properties. 
Flow fields of a small- and a medium-sized axisymmetric sunspot provide
the context for interpreting the results. All sunspots show the outward 
moat flow and the advection of moving magnetic features (MMFs). However, the 
extent of the moat flow varies from spot to spot but a correlation of flow
properties with size is tenuous, if at all present. The moat flow is 
asymmetric and predominantly in the east-west direction, whereby deviations
are related to the tilt angle of the sunspot group as well as to the 
topology and activity level of the trailing plage.} 


\jnlcitation{\cname{%
\author{M.\ Verma}, 
\author{P.\ Kummerow}, and 
\author{C.\ Denker}} (\cyear{2018}), 
\ctitle{On the extent of the moat flow in axisymmetric sunspots}, 
\cjournal{ASNA}, \cvol{2018;00:1--8}.}

\maketitle


\section*{Introduction}

Sunspots are surrounded by an annular horizontal flow field, the `moat flow',
which was first detected by \citet{Sheeley1972}. The moat flow was soon 
linked to sunspot decay \citep{Harvey1973}, where MMFs
\citep{Hagenaar2005} play an important role in transporting magnetic 
flux across the spot's boundary. Even residual pores, i.e., sunspots having 
lost their penumbrae, still exhibit MMFs and indications of moat flow
\citep{Deng2007, Zuccarello2009, Verma2012a}. However, the congruity between 
moat flow and sunspot decay is challenged by its absence around many pores
\citep{CabreraSolana2006, VargasDominguez2007} and its presence in earlier 
stages of sunspot evolution \citep{Brickhouse1988}. Thus, even four decades
after discovery, fundamental questions remain open regarding the moat flow's
origin and its relation to the spot's evolutionary stage, size, and penumbral
dimensions. 

Another outward radial flow starting at the sunspot's penumbra is the Evershed 
flow \citep{Evershed1909}. The Evershed flow is correlated with the 
inclination angle and horizontal strength of the penumbral magnetic field
\citep[see][for a sample of nine sunspots]{Deng2011a}. A similar correlation 
holds for averaged quantities including the total magnetic field strength 
with respect to the width of the penumbra. The relation between the Evershed 
flow and MMFs was studied by \citet{CabreraSolana2006} who analyzed
spectropolarimetric measurements of a sunspot in the Fe\,\textsc{i}
$\lambda$630.2~nm and $\lambda$1565~nm lines. They followed the temporal 
evolution of radially outward moving Evershed clouds along the same penumbral
filament. They concluded that the extension of the penumbral Evershed flow may 
appear as MMFs in the moat flow. \citet{Rempel2011c} explained 
the relationship between moat and Evershed flow by suggesting two
components of penumbral flows, one (deep) related to the moat flow and 
another (shallow) carrying the Evershed flow.

\citet{LoehnerBoettcher2013} used HMI Doppler maps to study moat and Evershed 
flow around 31 sunspots. From their analysis of three-hour averaged Doppler 
maps, they found that the moat flow velocity and its extent is independent 
of sunspot size. In contrast, they noticed an enhancement in the Evershed 
velocity with sunspot size. However, following the evolution of flows around
long-lived sunspots for six to eight days, they observed an increase in the 
moat flow velocity along with the decay of sunspots. 
In contrast, the Evershed flow velocity decreased. They concluded that
the moat flow is a non-magnetic flow deeper in the photosphere, whereas the 
Evershed flow is a magnetized flow extending into the sunspot canopy. They
regarded MMFs to be advected by the moat flow but they consider these two
phenomena to be different. Their view is in agreement with the work of
\citet{Nye1988} who used linear numerical models of mass and energy flow
suggesting that the driving force behind the moat flow is the stored thermal
energy blocked by the sunspot. They also postulated that the moat flow's 
extent is proportional to the depth of sunspot penumbra but not to the size
of sunspot.

\citet{Weiss2004} suggested that magnetic flux tubes are dragged downward
as a result of turbulent pumping by granular convection in the immediate
surroundings of the sunspot (the ``moat''), resulting in various types of MMFs.
\citet{Gafeira2014} presented another scenario, in which the velocity field
interacts with an axially symmetric and height-invariant magnetic field,
reproducing the large-scale features of the much more complex convection  
observed inside sunspots. Their results were in agreement with the Sun's
subsurface dynamics \citep{Kitiashvili2009}, where low-magnitude inflows 
of umbral and penumbral features can coexist with Evershed outflows as 
part of overturning convective motions.

\begin{table}[t]
\centering
\caption{Active region summary (area and radius).}
\footnotesize
\phn
\begin{tabular}{lccc}
\hline\hline
\textbf{NOAA}      & \textbf{11809}  & \textbf{12032} & \textbf{11899}\rule[-1.5mm]{0mm}{5mm}\\
\hline
Date               & 2013--08--06    & 2014--04--13   & 2013--11--18\rule[0mm]{0mm}{3.5mm}\\
Meridian passage   & 23:13~UT        & 18:06~UT       & 15:44~UT\\
$\mu=\cos\theta$   & 0.973           & 0.987          & 0.994\\
                   \cmidrule[0.4pt]{2-4}
                   &                 &  Area [Mm$^2$] & \rule[0mm]{0mm}{3.5mm}\\ 
                   \cmidrule[0.4pt]{2-4} 
$A_\mathrm{moat}$  &        1102   &     1600     &     5188 \\
$A_\mathrm{mag}$   &   \phn 241    & \phn 608     &     2231 \\
$A_\mathrm{spot}$  &   \phn 216    & \phn 626     &     2266 \\
$A_\mathrm{umbra}$ & \phn\phn 41   & \phn 128     & \phn 586 \\
                   \cmidrule[0.4pt]{2-4}
                   &                 &    Radius [Mm] & \rule[0mm]{0mm}{3.5mm}\\ 
                   \cmidrule[0.4pt]{2-4}
$r_\mathrm{moat}$  &    18.7         & 22.6           & 40.6\\ 
$r_\mathrm{mag}$   & \phn8.8         & 13.9           & 26.7\\ 
$r_\mathrm{spot}$  & \phn8.3         & 14.1           & 26.9\\ 
$r_\mathrm{umbra}$ & \phn3.6         & \phn6.4        & 13.7 \rule[-1.5mm]{0mm}{3mm}\\
\hline
\end{tabular}
\label{TAB01}
\end{table}

Based on EUV $\lambda$170~nm and continuum images of the Transition Region and Coronal 
Explorer \citep[TRACE,][]{Handy1999}, \citet{Balthasar2010} inferred the 
height dependence of the moat flow. In the upper atmosphere, flow speeds 
exceeded photospheric values in the inner moat while falling behind in the 
outer moat, which was attributed to a decoupling of the EUV bright-points 
from the photospheric plasma motions. However, \citet{Sobotka2007} observed 
that the moat areas are correlated in white-light and EUV images,
but moat widths are independent of the spot radius 
\citep[cf.,][]{Brickhouse1988}. In
addition, an East-West asymmetry was observed, i.e., in white-light 
images the moat extended more in towards the East. However, this trend is 
opposite for the moat as seen in the EUV images.


\section{Observations and data reduction}

On 2013 November~18, a large unipolar, axisymmetric sunspot crossed the central 
meridian of the Sun. Its exceptional size raises the question, if its horizontal 
flow field shows any peculiarities as compared to small- or intermediate-sized 
spots. We present preliminary results of an ongoing statistical study by 
focusing on three exemplary sunspots observed in 2013 and 2014 (see 
Table~\ref{TAB01}\ns). The SDO data \citep{Pesnell2012} consist of HMI 
continuum images and magnetograms \citep{Scherrer2012, Schou2012} with a
cadence of 45~s and AIA EUV $\lambda$160~nm images with 24-second sampling
\citep{Lemen2012}, i.e., half of the fastest cadence of 12~s. Each image 
or magnetogram of the 12-hour time-series is compensated for differential rotation using 
the meridian passage of the spot as time reference. A region-of-interest
(ROI) is extracted from the full-disk data, which is corrected for limb 
darkening (only continuum and EUV images) and geometric foreshortening. Thus,
the intensity $I$ is normalized to the local intensity of the quiet Sun, i.e.,
$I_\mathrm{con} / I_0$ and $I_\mathrm{EUV} / I_0$. The cross-correlation between
single continuum images and their 12-hour average identifies residual drifts, 
associated with active-region evolution, which are subsequently removed 
from all time-series. Resampling on a regular grid with a spacing of 0.4~Mm 
yields an ROI with dimensions of 160~Mm $\times$ 160~Mm. The data processing
essentially follows the procedures elaborated in \citet{Kummerow2015},
\citet{Verma2011}, and \citet{Beauregard2012}.

Continuum and EUV images are subjected to the LCT algorithm outlined in
\citet{Verma2011}, which is a variant of the technique put forward by
\citet{November1988}. A subsonic filter with a cut-off at the photospheric 
sound speed suppresses intensity variations caused by solar oscillations. 
Sampling window and high-pass filter are identical and have a size of 6.4~Mm
$\times$ 6.4~Mm. They are implemented as two-dimensional Gaussians with a 
full-width-at-half-maximum of 2~Mm. 
The cadence for correlating image pairs is $\Delta t = 90$~s and 96~s for 
continuum and EUV images, respectively. Individual flow maps are averaged 
over $\Delta T = 12$~hours to derive the persistent flows in and around 
sunspots.

Flux transport velocities are derived from magnetograms using DAVE 
\citep{Schuck2005, Schuck2006} with a 11$\times$11-pixel (4.4~Mm $\times$ 
4.4~Mm) sampling window. We coarsely assume that the magnetic field lines 
are perpendicular to the solar surface, thus, dividing the field strength
$B_\mathrm{LOS}$ along the line-of-sight (LOS) by the cosine of the 
heliocentric angle $\mu$ yields $B = B_\mathrm{LOS} / \mu$. 
Convolving the magnetograms with the appropriate \citet{Scharr2007} 
operators results in the spatial derivatives of the magnetic field in
the $x$- and $y$-directions, whereas the temporal derivative is based 
on a a five-point stencil with a 15-minute time interval. Taking into account
the 30-minute offsets at the beginning and end of the time-series because
of the five-point stencil, 880 individual flow maps enter the computation
of the time-averaged flow maps.


\section{Results}

The three sunspots are chosen as representative examples for small, medium, 
and large sunspots. Morphological image processing and thresholding provide
areas $A$ enclosed by the spot's photometric, magnetic, and moat boundaries.
Equivalent radii $r$ refer to circles that conform to the respective 
perimeters. To measure the sunspot's area, the average continuum image is
smoothed using a Perona-Malik filter \citep{Perona1990}, and fixed intensity
thresholds of $I_\mathrm{con} / I_{0} = 0.6$ and $I_\mathrm{con} / I_{0} =
0.92$ are applied to determine the extent of umbra and penumbra, respectively. 
The area and radius are then estimated using the blob-analyzer code by \citet{Fanning2003} as
described in \citet{Verma2014}. Magnetic area and radius are computed by
thresholding the magnitude of the magnetograms at $B = 100$~G. The areal 
extent of the moat is computed using DAVE velocity maps, where morphological smoothing
is applied to the radial component of the velocity, along with a threshold 
of 100~m~s$^{-1}$ for the radial velocity. We compare the moat's extend 
based on radial velocity maps for continuum and EUV images, but only the 
results from magnetograms led to a robust estimate for the moat's area. 
Area and radius for the three sunspots are listed in Table~\ref{TAB01}\ns.
In the remainder of this study, the physical parameters characterizing the flow fields
refer to the feature definitions above.

\begin{figure*}
\includegraphics[width=\textwidth]{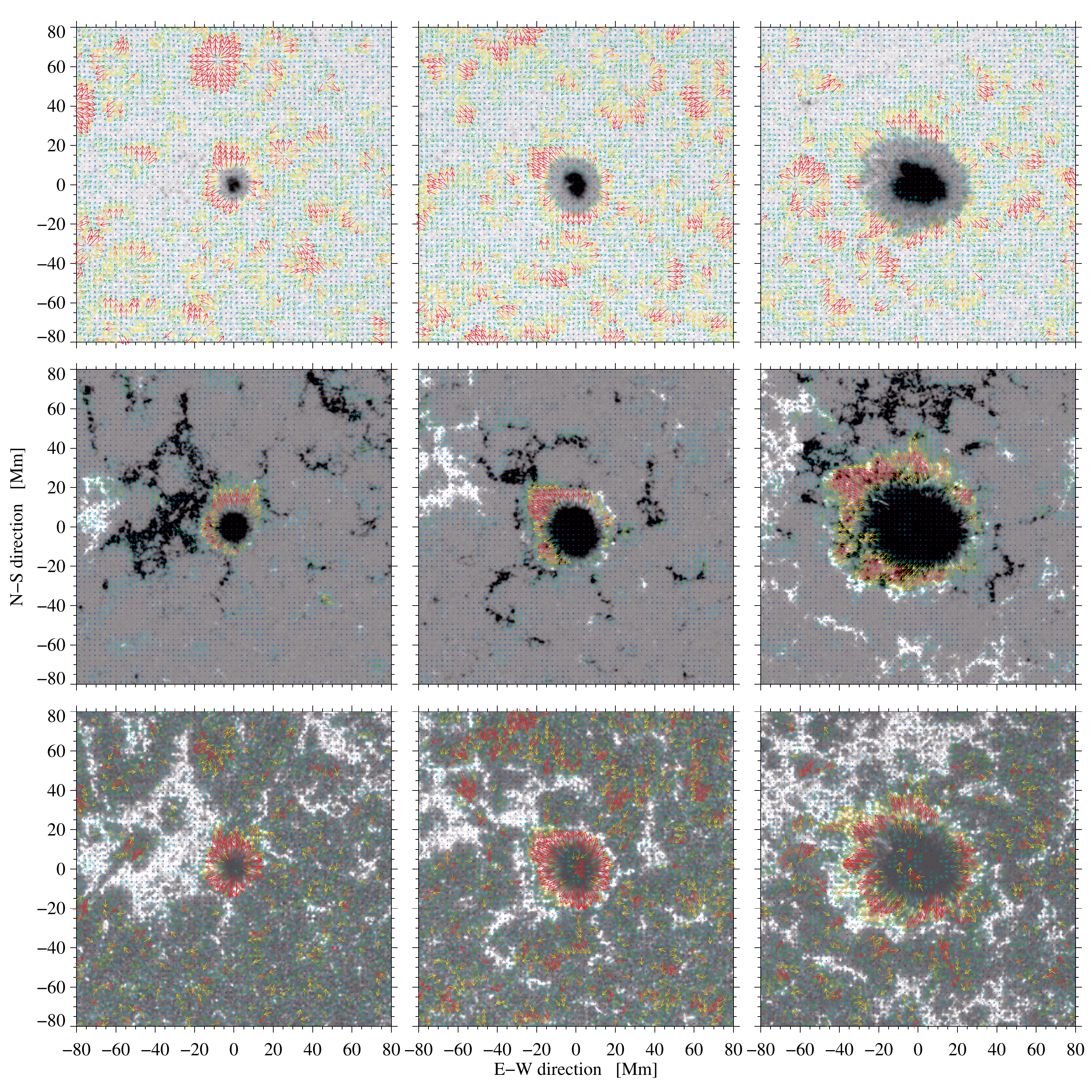}
\caption{Continuum intensity $I_\mathrm{con} / I_0$, magnetic flux density 
    $B = B_\mathrm{LOS} / \mu$, and EUV intensity $I_\mathrm{EUV} 
    / I_0$ (\textit{top to bottom}) for three axisymmetric sunspots in 
    active regions NOAA~11809, 12032, and 11899 (\textit{left to right}),
    respectively. The superimposed, rainbow-colored vectors represent 
    magnitude and direction of the horizontal proper motions. Violet and 
    red correspond in the top panels to flow speeds of lower than 0.05 and 
    larger than 0.25~km~s$^{-1}$, respectively. Scaling this range by 
    factors of 1.6 and 2.0, respectively, accommodates the higher flow 
    speeds derived from magnetograms and EUV images.}
\label{FIG01}
\end{figure*}

Maps depicting time-averaged horizontal proper motions and flux transport velocities 
are compiled in Fig.~\ref{FIG01}\ns\ for active regions 
NOAA~11809, 12032, and 11899, which
exemplarily displays a small, a medium, and a large sunspot, respectively. 
The background images are continuum images, EUV images, and magnetograms 
at meridian passage. Flow vectors exist for each of the $400 \times 400$ pixels in the FOV but for 
display purposes the flow vectors are resampled onto an equidistant grid 
of $60 \times 60$ points. However, quantitative flow speeds always refer 
to averages over all pixels belonging to a certain feature.

Spatial resolution and contrast of continuum images are sufficient 
to measure persistent flows in the granulation. Thus, the continuum flow
maps contain additionally signatures of meso- and supergranular flows
\citep{November1988}. Furthermore, prominent circles with close to zero 
flow speeds \citep{Molowny-Horas1994a, Denker1998b, Deng2007} are visible
in the middle of the penumbra (divergence) and at the termination of the
moat flow (convergence). Even though we select unipolar, axisymmetric
sunspots, the flow fields exhibit significant asymmetries \citep[see
e.g.,][]{Sobotka2007}. For all three sunspots the moat flows are lower on the
south-west side.

The velocities patterns seen in the flow maps based on magnetograms are 
virtually identical to those based on continuum images, except the 
pronounced meso- and supergranular pattern is absent, and the flow field around 
the sunspot is more cohesive. Therefore, the moat
flow is much easier discernible in these maps depicting flux transport
velocities, and the aforementioned asymmetries are also present. In EUV
images, the contrast of bright-points belonging to the chromospheric 
network and decaying plage is even higher -- exceeding the granular 
contrast by far. The flow maps based on EUV images show that the horizontal
proper motions are suppressed in the plage regions, i.e., at the location 
of dispersed small-scale magnetic flux elements. However, taking into account the
entire quiet Sun, the flow speeds are highest for the EUV flow maps. Furthermore,
the moat flow is more symmetric in EUV flow maps as compared to flow maps based 
on continuum images and magnetograms.

The photometric and magnetic radii are virtually identical. These radii 
are about 8.5, 14, and 27~Mm for the three sunspots,
whereas the areas are about 230, 620, and 2250~Mm$^2$. The respective moat areas 
are about 1100, 1600, and 5200~Mm$^2$. These values show that the moat area is 
about two times larger than the spot area for the medium and large sunspot, 
whereas the moat area is almost five times larger for the small sunspot.
Expressing these findings in terms of the radius, the ratios of moat 
to spot radius $r_\mathrm{moat} / r_\mathrm{spot}$ are 2.3, 1.6,
and 1.5 for the small, medium, and large spot, respectively.

To quantify the flow properties, we compile the mean flow speeds for 
moat, penumbra, umbra, and quiet Sun in Table~\ref{TAB02}\ns. The mean 
flux transport velocities are higher in the moat for all three spots, 
followed closely by the horizontal proper motions derived from EUV images. 
The flow speeds for penumbra and umbra are virtually identical for flow maps
based on continuum images and magnetograms, whereas they are higher for maps
based on EUV images. The quiet-Sun flow speeds for continuum and EUV maps 
are of the same order while the flux transport velocities are smaller. 
A likely explanation is that DAVE does not track very faint, small-scale
magnetic elements in the quiet Sun. Interestingly, the flow speed in 
the moat is highest for the medium-sized sunspot and lowest for the 
large sunspot in all three types of flow maps. Although, this discrepancy 
is most evident in the flow speeds and flux transport velocities
derived from continuum images and magnetograms, respectively.

\begin{table}[t]
\centering
\caption{Active region summary (horizontal proper motions and flux transport velocities).}
\footnotesize
\phn
\begin{tabular}{lccc}
\hline\hline
\textbf{Features}                           & \textbf{Continuum}  & \textbf{Magnetogram} & \textbf{EUV}\rule[-1.5mm]{0mm}{5mm} \\
                                            & $\bar{v} \pm \sigma_{v}$ [m~s$^{-1}$] 
                                            & $\bar{v} \pm \sigma_{v}$ [m~s$^{-1}$] 
                                            & $\bar{v} \pm \sigma_{v}$ [m~s$^{-1}$] \rule[-1.5mm]{0mm}{3mm} \\ 
\hline
                                            &                     &     \bf{11809}       & \rule[0mm]{0mm}{3.5mm} \\
                                            
                                            \cmidrule[0.4pt]{2-4}
Moat                                        &    165 $\pm$    117 &    277 $\pm$    \phn94 & 221 $\pm$    152 \\
Penumbra                                    & \phn66 $\pm$ \phn56 & \phn45 $\pm$    \phn27 & 166 $\pm$    168 \\
Umbra                                       & \phn24 $\pm$ \phn20 & \phn34 $\pm$ \phn\phn7 & 130 $\pm$ \phn90 \\
Quiet Sun                                   &    142 $\pm$    102 & \phn82 $\pm$    \phn47 & 162 $\pm$    129 \\
                                            \cmidrule[0.4pt]{2-4}
                                            &                     &     \bf{12032}         & \rule[0mm]{0mm}{3.5mm} \\ 
                                            \cmidrule[0.4pt]{2-4} 
Moat                                        &    219 $\pm$    123 &    312 $\pm$       109 & 267 $\pm$ 155 \\
Penumbra                                    & \phn95 $\pm$ \phn71 & \phn97 $\pm$    \phn79 & 228 $\pm$ 188 \\
Umbra                                       & \phn46 $\pm$ \phn35 & \phn41 $\pm$    \phn23 & 279 $\pm$ 204 \\
Quiet Sun                                   &    149 $\pm$    104 & \phn75 $\pm$    \phn44 & 175 $\pm$ 132 \\                                                               
                                            \cmidrule[0.4pt]{2-4}
                                            &                     &     \bf{11899}         &  \rule[0mm]{0mm}{3.5mm} \\ 
                                            \cmidrule[0.4pt]{2-4} 
Moat                                        &    131 $\pm$    101 &    221 $\pm$       132 & 179 $\pm$ 122 \\
Penumbra                                    & \phn90 $\pm$ \phn75 &    165 $\pm$       109 & 182 $\pm$ 147 \\
Umbra                                       & \phn50 $\pm$ \phn41 & \phn54 $\pm$    \phn27 & 167 $\pm$ 132 \\
Quiet Sun                                   &    132 $\pm$ \phn91 & \phn77 $\pm$    \phn46 & 150 $\pm$ 126 \\                 

\hline                                            
\end{tabular}
\label{TAB02}
\end{table}







\begin{figure*}[t]
\includegraphics[width=\textwidth]{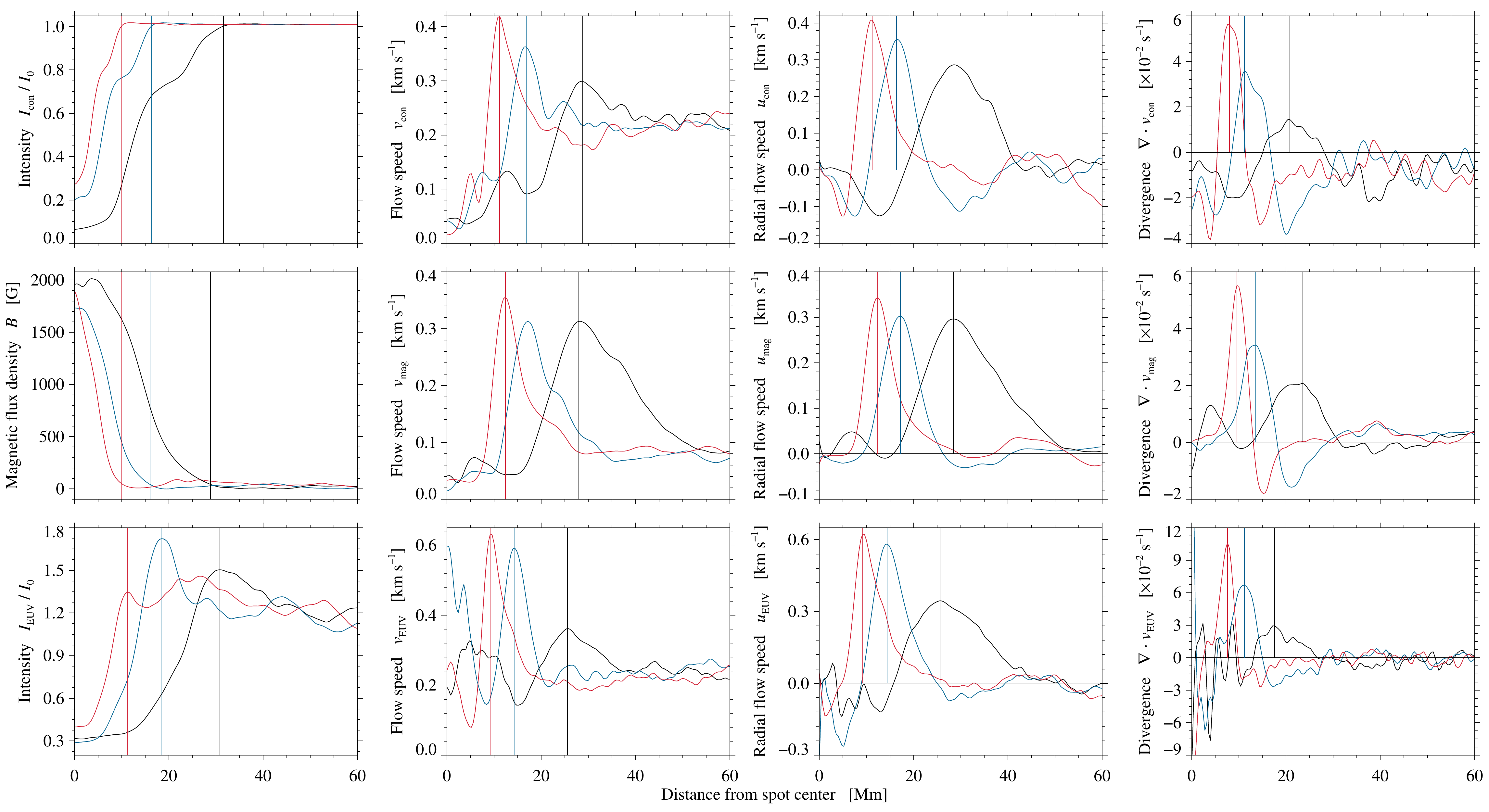}
\caption{Radial averages of intensity $I$, magnetic flux density $B$, 
   horizontal flow speed $v$, radial component of the flow speed $u$, and 
   divergence $\nabla \cdot v$ (\textit{left to right}). The profiles are 
   based on continuum images, magnetograms, and EUV 
   images (\textit{top to bottom}). Red, blue, and black 
   lines refer to active regions NOAA~11809, 12032, and 11899, respectively. 
   The vertical lines mark characteristic parameters of the displayed radial profiles
   and the corresponding radial distances from the center of the spots, which
   are summarized in Table~\ref{TAB03}\ns.}
\label{FIG02}
\end{figure*}

Radial averages of sunspot properties facilitate an easier comparison of
sunspots with different sizes. In this sense, Fig.~\ref{FIG02}\ns\ complements 
the two-dimensional maps of Fig.~\ref{FIG01}\ns, and we introduce additional
parameters describing the flow field. The horizontal velocity vector 
$\vec{v} = (v_x, v_y)$ is decomposed yielding the radial component of the 
flow velocity $u = v_x \cos\varphi + v_y \sin \varphi$, where $\varphi$ 
is the azimuth about the spot's center measured counterclockwise from west
(positive $x$-axis). The divergence $\nabla\! \cdot \vec{v} = 
\partial v_x / \partial x + \partial v_y / \partial y$ is computed using 
finite differences. Table~\ref{TAB03}\ns\ lists the characteristic
values for the averaged radial profiles of various physical parameters shown in 
Fig.~\ref{FIG02}\ns\ along with the corresponding radial distance from the
center of the spots.


\begin{table}[t]
\centering
\caption{Active region summary (radial distance to maximum values).}
\footnotesize
\phn
\begin{tabular}{lccc}
\hline\hline
\textbf{Parameters}                         & \textbf{Continuum} & \textbf{Magnetogram} & \textbf{EUV }\rule[-1.5mm]{0mm}{5mm}\\
\hline
                                            &                    &     \bf{11809}       &  \rule[0mm]{0mm}{3.5mm}\\
                                            \cmidrule[0.4pt]{2-4}
$I_\mathrm{max} / I_0$, $B_\mathrm{max}$ [G] &  1.0              & 50                   & 1.7   \\
                                             &  10.0             & 10.0                 & 11.2\rule[-2mm]{0mm}{3mm}  \\            
$v$ [km~s$^{-1}$]                            &  0.42             & 0.36                 & 0.63  \\
                                             &  11.2             & 12.4                 & 9.2\rule[-2mm]{0mm}{3mm}   \\
$u_\mathrm{rad}$ [km~s$^{-1}$]               &  0.42             & 0.34                 & 0.62  \\
                                             &  11.2             & 12.4                 & 9.2\rule[-2mm]{0mm}{3mm}   \\
$\nabla\! \cdot \vec{v}$ [$\times 10^{-2}$~s$^{-1}$] &  0.056    & 0.046                & 0.10  \\
                                             &  8.0              & 9.6                  & 7.6   \\                                          
                                            \cmidrule[0.4pt]{2-4}
                                            &                    &     \bf{12032}        &  \rule[0mm]{0mm}{3.5mm}\\ 
                                            \cmidrule[0.4pt]{2-4} 
$I_\mathrm{max} / I_0$, $B_\mathrm{max}$ [G] &  1.0              & 50                   & 1.7   \\
                                             &  16.4             & 16.0                 & 18.4\rule[-2mm]{0mm}{3mm}  \\            
$v_\mathrm{rad}$ [km~s$^{-1}$]               &  0.36             & 0.31                 & 0.59  \\
                                             &  16.8             & 17.2                 & 14.4\rule[-2mm]{0mm}{3mm}  \\
$u_\mathrm{rad}$ [km~s$^{-1}$]               &  0.36             & 0.30                 & 0.58  \\
                                             &  16.4             & 17.2                 & 14.4\rule[-2mm]{0mm}{3mm}  \\
$\nabla\! \cdot \vec{v}$ [$\times 10^{-2}$~s$^{-1}$ &  0.038     & 0.027                & 0.067 \\
                                             &  11.2             & 13.6                 & 11.2  \\                          
                                            \cmidrule[0.4pt]{2-4}
                                            &                    &     \bf{11899}       &  \rule[0mm]{0mm}{3.5mm}\\ 
                                            \cmidrule[0.4pt]{2-4}
$I_\mathrm{max} / I_0$, $B_\mathrm{max}$ [G] &  1.0              & 50                   & 1.5   \\
\rule{23mm}{0mm}                             &  31.6             & 28.8                 & 30.8\rule[-2mm]{0mm}{3mm}  \\            
$v_\mathrm{rad}$ [km~s$^{-1}$]               &  0.30             & 0.31                 & 0.36  \\
                                             &  28.8             & 28.0                 & 25.6\rule[-2mm]{0mm}{3mm}  \\
$u_\mathrm{rad}$ [km~s$^{-1}$]               &  0.29             & 0.30                 & 0.34  \\
                                             &  28.8             & 28.4                 & 25.6\rule[-2mm]{0mm}{3mm}  \\
$\nabla\! \cdot \vec{v}$ [$\times 10^{-2}$~s$^{-1}$] &  0.019    & 0.016                & 0.029 \\
                                             &  20.8             & 23.6                 & 17.6  \\   
\hline                                             
\end{tabular}
\footnotesize
\hspace*{0.5mm}
\noindent\parbox{74mm}{\vspace*{1mm}
\begin{itemize}
\item[Note:] $v$ is the flow speed, $u_\mathrm{rad}$ is the radial flow speed, 
and $\nabla\! \cdot \vec{v}$ is the divergence. The values below the physical 
parameters refer to the radial distance in megameters.
\end{itemize}}
\label{TAB03}
\end{table}

Inspecting radial profiles for axisymmetric sunspots motivates different 
definitions of thresholds for features using inflection points and extreme values.
The continuum radius $I_\mathrm{con} / I_{0} = 1$ and magnetic radius
$B=50$~G are virtually identical with about 10, 16, and 32~Mm 
for the small, medium, and large sunspot, respectively. The radii derived 
from azimuthal averages are somewhat larger than those based on thresholds for
two-dimensional maps of intensity and magnetic field. The normalized EUV 
intensity has two peaks.
One at the location matching the continuum and magnetic radii, and a 
secondary peak further out. The first peak is the prominent one for 
the medium and large sunspot, but not for the small sunspot, where the
secondary peak is stronger. The secondary peaks are mainly 
caused by EUV bright-points. The flow speed based 
on continuum images and magnetograms peaks at the location of the continuum 
and magnetic boundaries for all sunspots. In the EUV, we
find a peak just before the continuum and magnetic boundaries.

Azimuthal averages of the radial flow speed have a similar trend for 
horizontal proper motions based on the continuum images for all sunspots. 
In the spot's center the radial flow speed is zero, but starts to decrease
(negative speed indicates inflows) reaching a minimum at around the radial
distance that corresponds to $I_\mathrm{con} / I_{0} \approx 0.6$. 
The radial flow speed returns to zero inside the penumbra at
$I_\mathrm{con} / I_{0} \approx 0.7$. The negative radial flow speed is caused
by inward migrating penumbral grains with high contrasts and may not reflect a 
real plasma motion. The maximum of the radial flow 
speed is reached $I_\mathrm{con} / I_{0} \approx 1$. Moreover, the 
radial flow speeds based on magnetograms and EUV images follow a similar 
trend deviating only inside the spots, where the radial flow speeds are 
much noisier.

In the continuum, the divergence is negative inside the three sunspots, 
similar to the radial flow speed. The divergence switches to positive value 
inside the sunspots before attaining a maximum. The location, where the 
divergence reaches its maximum, is inside the spot's penumbra. Once the 
maximum is reached the divergence slowly approaches zero. Outside sunspots the 
divergence fluctuates between small positive and negative 
values.

\begin{SCfigure*}[0.7065][t]
\includegraphics[width=130mm]{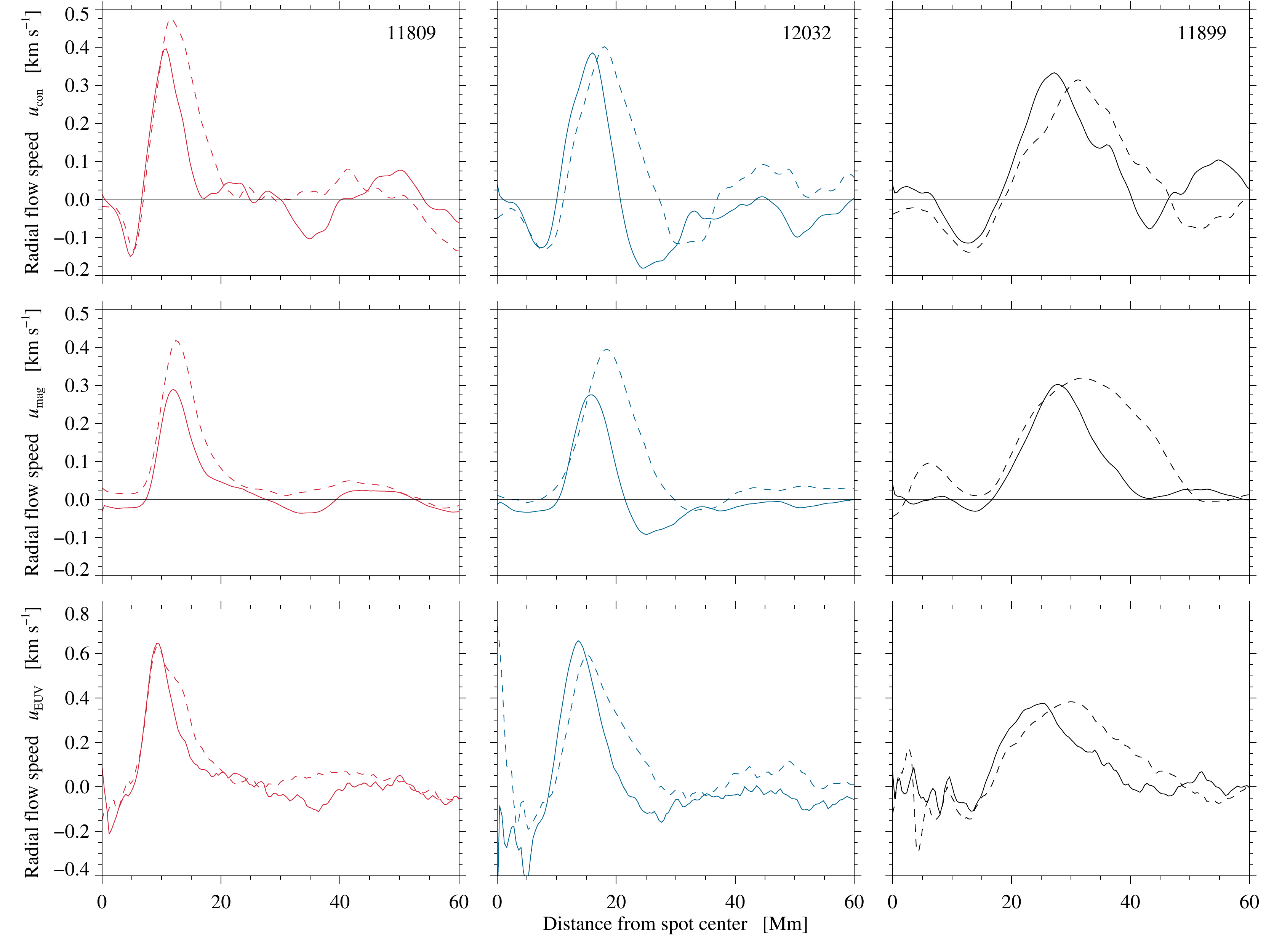}
\hspace*{2mm}
\caption{The east (\textit{dashed}) and west (\textit{solid}) components of 
   radial velocities based on continuum images,  magnetograms, 
   and EUV images (\textit{top to bottom}). Red, blue, and 
   black lines refer to active regions NOAA~11809, 12032, and 11899, respectively.}
\label{FIG03}
\end{SCfigure*}

To quantitatively assess asymmetries in the moat flow, we plot the radial
velocity separately for East and West sides of the three sunspots and the
three data types in Fig.~\ref{FIG03}\ns. Asymmetric flows appear for the 
east component as broader profiles with maxima shifted further out. However, 
the width of the profiles is the distinguishing characteristic, whereas the
displacement of the maxima is less pronounced. The largest displacement of
about 8~Mm is observed for the largest sunspot. In most cases, the values of 
maxima are virtually identical for both sides, except for the small and medium
sunspots, where the east side exhibits higher values for radial continuum and
flux transport velocities. This East-West asymmetry in the moat flow was also
observed by \citet{Kummerow2015} in the averaged flow field maps of 26 
sunspots, corroborating the findings of \citet{Sobotka2007}.

\begin{figure*}
\includegraphics[width=\textwidth]{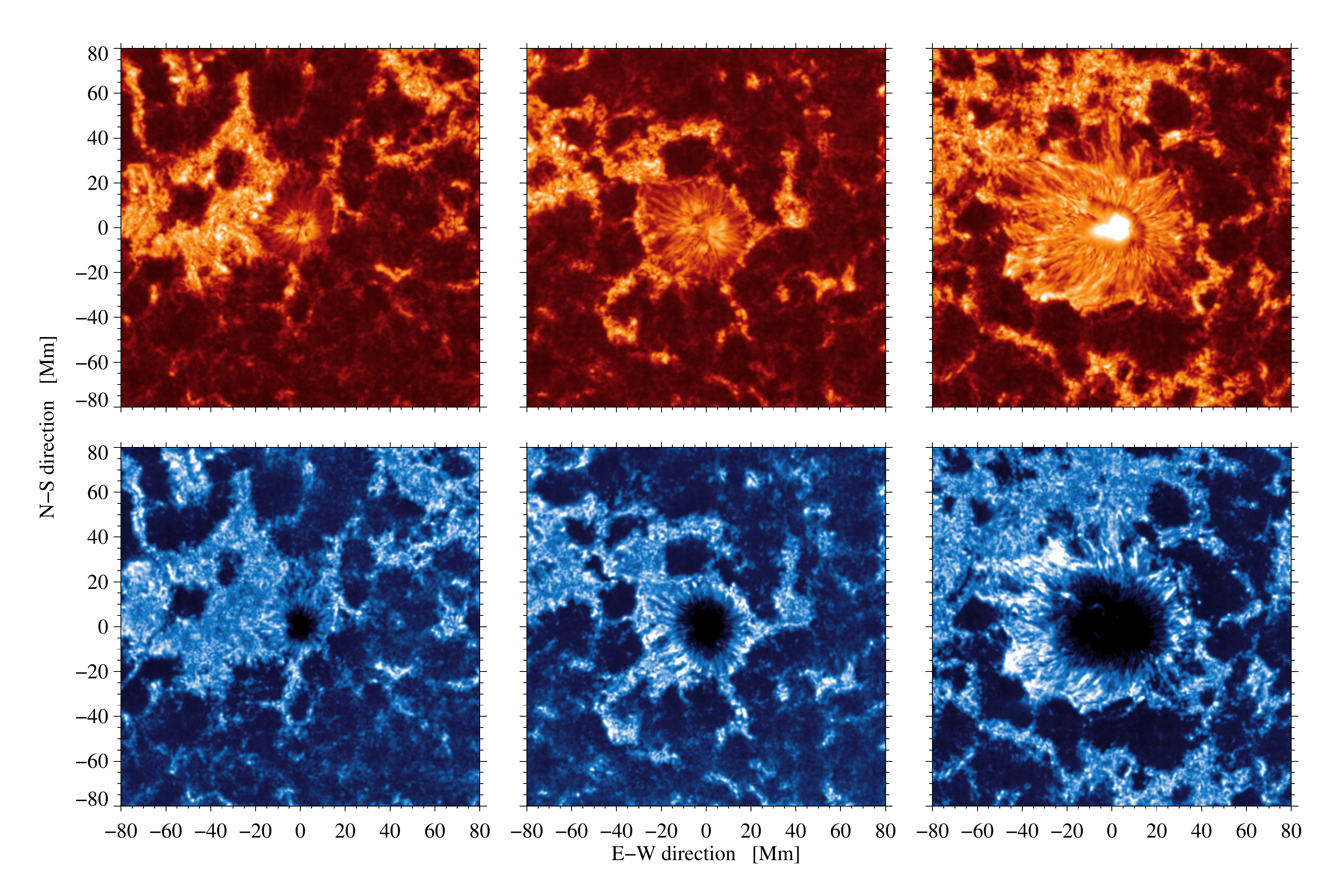}
\hspace*{2mm}
\caption{Background-subtracted variations of the magnetic flux density 
    (\textit{top}) $\langle\,|\,B-\langle B\rangle\,|\,\rangle$ and the EUV
    intensity (\textit{bottom}) $\langle\,|\,I-\langle I\rangle\,|\,\rangle$ 
    in active regions NOAA~11809, 12032, and 11899 (\textit{left to right}),
    where $\langle\,\cdots\rangle$ indicates an ensemble average (logarithmic
    display).}
\label{FIG04}
\end{figure*}

For all three sunspots, we plot the background-subtracted variation of 
the magnetic flux density $\langle\,|\,B-\langle B\rangle\,|\,\rangle$ and 
the EUV intensity $\langle\,|\,I-\langle I\rangle\,|\,\rangle$
(Fig.~\ref{FIG04}\ns). In all three cases, the sunspots occupy the center 
of a supergranular cell. Spoke-like structures join the sunspots to the
surrounding supergranular network, as is evident in the variation of 
magnetic flux density and EUV intensity. This kind of radial structure 
was first reported by \citet{Verma2012a} around a decaying sunspot. The 
length of these spokes varies for all three sunspots. They are longest for 
the large sunspot, where the length of the spokes varies between 15--20~Mm.
An east-west asymmetry is present in the medium and large sunspot, 
with longer spokes extending towards the east. The variation maps are a 
versatile tool to visualize ongoing changes in active regions. They also 
clearly demonstrate the connectivity of horizontal proper motions and flux
transport velocities within the entire active region.


\section{Discussion and conclusions}

Three axisymmetric sunspots with different radii were selected to 
scrutinize horizontal proper motions based on continuum and EUV images
(LCT) and flux transport velocities based on magnetograms (DAVE). 
Our study complements the work by \citet{LoehnerBoettcher2013}, 
which was based on HMI Doppler maps. In proximity to the solar limb, 
horizontal flows in and around sunspots can be determined by decomposing 
the Doppler velocity. In contrast, our measurements yield horizontal flows
in and around sunspots in proximity to the disk center.
\citet{LoehnerBoettcher2013} found that the moat flow size and velocities
are independent of the sunspot size, similar to previous work by
\citet{Sobotka2007}. In addition, the former authors concluded that MMFs are
related to the inclined magnetic field lines (relative to horizontal),
and although they are advected by the moat flow, they are distinguishable from it. 
In the present work, the moat radius of the small sunspot is 2.25 times 
the photometric radius, whereas for the medium and large sunspot it is 
1.6 and 1.5 times the photometric radius, respectively. Hence, we also 
do not find any clear relation between the photometric and the moat 
radius. Even the much larger sample of 26 sunspots scrutinized by 
\citet{Kummerow2015} shows only a tenuous correlation between
sunspot and moat radius, if at all.

Large, isolated, axisymmetric sunspots are more likely to emerge at 
low latitudes during the declining phase of the solar cycle
\citep{Koutchmy2009}. The large sunspot of this study was chosen as 
an example from the declining phase of solar cycle No.~24. A medium 
and a small sunspot were used for comparison. The horizontal flow 
properties in the continuum were similar for all three spots, with 
an outflow encircling the spots and inflows in the inner penumbra. 
The basic properties of the moat flow remained the same irrespective 
of the spot's size. An east-west asymmetry of moat flow was observed 
for all spots, which however was more  prominent in the flow maps 
based on continuum images and magnetograms. This result is in agreement
with the work of \citet{Sobotka2007}. However, in a case study,
\citet{Balthasar2009} did not find any asymmetry in the moat flow, 
when a sunspot was observed in both white-light and EUV images.

The asymmetric moat flow is not strictly oriented in the east-west 
direction. Thus, the tilt angle of the active region and the distribution
of the neighboring plage may significantly impact the orientation. The overall
topology of leading axisymmetric sunspots and trailing plage regions may
also affect the height-dependent moat flow observed in continuum and EUV flow maps.
Moreover, the maps of the EUV intensity and magnetic field variation show 
that the moat region is most ``active'' for the large spot, whereas the 
plage regions of the small and medium spots show enhanced activity. Is this 
an indication that these spots are stable and that the evolution of the active
region is governed by the trailing plage? The strongest EUV variations are 
seen for the large sunspot on the eastern side. Here, the spot is most closely
connected to the trailing plage. Addressing these issues will be deferred to a
forthcoming more comprehensive statistical study encompassing all available 
SDO data.

The EUV continuum ($\lambda$160~nm) forms just below the temperature
minimum about 450--500~km above the photosphere \citep{Vernazza1981}. 
Hence, computing proper motions based on EUV images gives access to 
higher atmospheric layers. This approach has already been followed by
\citet{Sobotka2007} and \citet{Balthasar2009}, when they used TRACE 
white-light and EUV images. In the azimuthally averaged intensity 
around the three sunspots in the current study, a peak appears in EUV 
similar to what was observed by \citet{Balthasar2009}. This indicates 
the presence of long-lived, small-scale bright-points around sunspots, 
which appear higher in the atmosphere. The moat flow itself was well 
traceable in the EUV  images. Flux transport velocities were computed 
with DAVE, which match the horizontal proper motions derived from 
continuum images. However, the typical inward motion in the inner 
penumbra is absent in the flux transport velocities. A likely explanation 
is the inward motion of penumbral grains visible in time-series of 
continuum images. These can be interpreted as the footpoints of flux
tubes, which become more vertical so that the bright footpoints migrate
inwards \citep{Schlichenmaier1998a, Schlichenmaier1998b}. Such a phase 
velocity resulting from systematic contrast changes will not be picked 
up by DAVE.

\citet{Svanda2014} applied time-distance helioseismic inversions to 
study the properties of sub-surface moat flows in 104 H-class sunspots and over 
200\,000 supergranule cells. Even though, they found similarities in both flows, 
and the moat flows replace the surpergranular flows around H-class sunspot, there 
are two major differences. First, the moat flow is asymmetric due to the 
sunspot motion across the solar disk and is disturbed by the surrounding plasma 
resulting in a larger extension on the eastern side than on the western side. 
In contrast, the outflow region in supergranules is symmetric. Second, studying the vertical 
flows, they noticed that the moat regions exhibit exclusively downflows, whereas 
supergranules have upflows in the center which turn into downflows at 60\% 
cell radius. This is related to mass sinking down around the H-sunspot in a
shallow sub-surface layer, which is at least twice the mass submerging on average 
in supergranular cells.

The horizontal flow properties around pores were summarized by 
\citet{Verma2014}. The radial average of the horizontal velocities 
revealed that pores, i.e., spots which never developed a penumbra, 
have a moat-like outflow structure encircling them. Moat flows are also
seen around decaying sunspots \citep[e.g.,][]{Balthasar2013, Verma2012a} 
-- even when the sunspot shrunk to become a residual pore, i.e., 
loosing its penumbra. This raises the question, is a penumbra needed 
for setting up moat flows or is the moat flow a part of an even larger 
flux system? In a recent study, \citet{Strecker2015} raised the question
if the moat flow is embedded in a cell of the supergranular network. 
She employed the method described by \citet{LoehnerBoettcher2013} to
compute horizontal proper motions for eight sunspots from limb-to-limb.
By following the evolution of horizontal velocity she concluded that
there is a constant interaction between moat flows and the surrounding
supergranular cells at all stages of sunspot evolution.
Maps of the EUV intensity and the magnetic field variation exhibit EUV
bright-points and magnetic features, which migrate from all three sunspots
towards the surrounding network. These easily perceived, radial paths are
followed by the MMFs and overlap with the encircling moat flow.
\citet{Verma2012a} already identified these radial spokes around a 
decaying sunspot. However, in the present study, none of the sunspots
were decaying but they all were in a rather mature state. This clearly
demonstrates that sunspots are part of a larger magnetic flux system, 
which encompasses an entire supergranular cell.

In recent work, \citet{Rempel2011a} and \citet{Rempel2015} investigated 
the relationship between moat and Evershed flow. Three magnetic components 
are involved in the flow patterns associated with sunspots: (1) the 
horizontal field leading to magnetized horizontal flows, (2) the vertical 
field that does not possess a horizontal flow component, and (3) the 
deep non-magnetized flows that predominantly contribute to the moat flow. 
In conclusion, the Evershed flow is a magnetized flow because of the
inclined magnetic fields and the overturning convection, whereas the 
moat flow results from a perturbation of the up- and downflow balance 
around the spot. In the present study, we confirm that certain flow 
properties are common for sunspots irrespective of their size, e.g., 
the outward moat flow and the advected MMFs. The properties of moat 
flow vary from spot to spot, but the overall trend remains the same. 
The MMFs moved to the surrounding network in all the three sunspots and 
are partly carried by the encircling moat. 

The uniformity of HMI and AIA data facilitates a quantitative 
comparison of sunspots, among others, of different size and evolutionary 
stage -- even statistical studies covering rise and fall of solar cycle 
No.~24 are now possible. We have presented numerical methods and analysis
tool, which allowed us to compare the flow properties of ``theoretican's 
sunspots'' from the smallest spots to one of the largest specimen. The
radial velocity profiles and the background-subtracted solar activity maps
for magnetograms and EUV intensity have proven very versatile in establishing
the connection between the sunspots and the surrounding supergranualar cells,
which deserve a holistic treatment in modeling.


\section*{Acknowledgments}

SDO HMI and AIA data are provided by the Joint Science Operations Center --
Science Data Processing. MV and CD acknowledge support 
by grant DE 787/5-1 of the German Science Foundation (DFG).



\end{document}